\documentclass[prd,showpacs]{revtex4}
\usepackage{graphicx}
\begin{document}
\title{
Thermal neutrino self-energy beyond the contact approximation}
\author{Andrea Erdas}
\email{aerdas@loyola.edu}
\affiliation{
 Department of Physics, Loyola University Maryland, 4501 N Charles Street,
       Baltimore, MD 21210}
\date{January 3, 2023}
\begin {abstract} 
Using the Schwinger proper time method I calculate exactly the bubble diagram contribution to the thermal neutrino self-energy, obtaining exact results that cannot be achieved using the contact approximation. These results allow me to obtain the dispersion relation for neutrino in a medium under several different conditions.
\end {abstract}
\maketitle
\section{Introduction}

Understanding accurately the propagation of neutrinos in a medium is important in particle physics, astrophysics, and cosmology~\cite{raffeltbook,giuntibook}. In the last thirty-some years a considerable effort has gone to theoretical studies of neutrino propagation and properties in a medium \cite{Wolfenstein:1977ue,Wolfenstein:1979ni,Mikheev:1986gs,Bethe:1986ej,Notzold:1987ik,Pal:1989xs,Langacker:1982ih,Langacker:1992xk,Enqvist:1990ad,DOlivo:1992lwg,Friedland:2003dv}, and in a medium with a magnetic field \cite{Erdas:1990gy,Erdas:1998uu,Erdas:2009zh,Borisov:1985ha,Borisov:1988wy,Borisov:1989yw,Elmfors:1996gy,DOlivo:1989ued,Kuznetsov:2005tq,Mckeon:1981ym,Elizalde:2000vz,Elizalde:2004mw,Nieves:2018qwg,Tarazona:2017jnd,Popov:2019nkr,Kuznetsov:2006ix}. All these studies do either a calculation of the neutrino self-energy in a medium by employing finite temperature field theory techniques, or a calculation of the self-energy in a magnetic field using exact propagators in magnetic field \cite{Erdas:1990gy}, or employ both techniques \cite{Erdas:1998uu} to investigate neutrino propagation and properties in a magnetized medium. However, none of those papers evaluates the non-local thermal part of the self-energy exactly, they all resort to the contact approximation calculation. In this paper, I will use Schwinger's proper time method \cite{Schwinger:1951nm} to calculate exactly the thermal corrections to the non-local part of the neutrino self-energy, i. e. the bubble diagrams. I will then use this result to calculate the neutrino dispersion relation in a variety of different media. I will investigate the case where a magnetic field is present, a very different calculation, in a future paper.

I introduce my notation in Sec. \ref{2}  and present the proper time method form of the fermion, vector, and scalar propagators in vacuum and in a medium. I use these propagators in Sec. \ref{3} to obtain exactly the non-local part of the thermal self-energy, i.e. the bubble diagram contribution, which is the main result of this paper. In Sec. \ref{4} I use this result to derive the thermal contribution of the bubble diagram in the case of low temperature and low neutrino energy. In Sec. \ref{5} I obtain the same quantity in the case of high temperature and high neutrino energy. Finally, in Sec. \ref{6}, I present the neutrino dispersion relation for all the cases examined before, and draw my conclusions.
\section{ Propagators in vacuum and in a medium}
\label{2}
In this paper I use the metric 
$g^{\mu \nu} = \mathrm{diag}(-1,+1,+1,+1)$ and work in the Feynman gauge, where the burden of keeping track of the contribution of unphysical scalars is a trade-off for a much simpler expression of the $W$-propagator. The
following expressions for the charged lepton
$S_0(k)$, $W$-boson
$G_0^{\mu \nu}(k)$, and scalar $D_0(k)$ vacuum propagators are written using Schwinger's proper time method  \cite{Schwinger:1951nm,Erdas:1990gy} :
\begin{equation}
\label{S0}
S_0(k) =i\int_0^\infty \!\!{ds} \,\,
{e^{-is\left(m^2_\ell+k^2\right)}}
(m_\ell-
\not\! k) , 
\end{equation}
\begin{equation}
G_0^{\mu \nu}(k)=
i\int_0^\infty \!\!{ds} \,\,
{e^{-is\left(m^2_W+k^2\right)}}
g^{\mu \nu},
\label{G0}
\end{equation}
\begin{equation}
D_0(k)=
i\int_0^\infty \!\!{ds} \,\,
{e^{-is\left(m^2_W+k^2\right)}},
\label{D0}
\end{equation}
where 
$m_\ell$ is the mass of the charged lepton $\ell$ and thus $\ell$ can take the values $e, \mu, \tau$,
$m_W$ is the $W$-mass, and the subscript $0$ on the propagators indicates explicitly that these are vacuum propagators. 

The thermal propagators of the real-time formalism of finite temperature field theory are constructed from
the proper-time form of the propagators in vacuum. The fermion 
propagator in a medium, $S(k)$, is written below in terms of the vacuum propagator $S_0(k)$ of Eq. (\ref{S0}) and the fermion
occupation number $f_F(k^0)$ 
\begin{equation}
S(k)=S_0(k)-f_F(k^0)\Bigl[S_0(k)-S_0^\ast (k)\Bigr]\,\,,
\label{S}
\end{equation}
where the piece proportional to $f_F$ is the thermal part of the propagator. The fermion occupation number at temperature $T$ and chemical potential
$\mu$ is defined as 
\begin{equation}
f_F(k^0)=f_F^+(k^0)\theta(k^0)+f_F^-(k^0)\theta(-k^0)
\label{fF}
\end{equation}
with
\begin{equation}
f_F^\pm(k^0)={1\over e^{\pm(k^0-\mu)/T}+1}\,\,.
\end{equation}
I obtain the thermal propagators of the $W$ boson, $G^{\mu\nu}(k)$, 
and of the charged scalar, $\Delta(k)$, in the same way:
\begin{equation}
G^{\mu\nu}(k)=G^{\mu\nu}_0(k)+f_B(k^0)\Bigl[G^{\mu\nu}_0(k)-
{G^{\mu\nu}_0}^\ast(k)\Bigr],
\label{G}
\end{equation}
and
\begin{equation}
\Delta(k)=\Delta_0(k)+f_B(k^0)\Bigl[\Delta_0(k)-\Delta_0^\ast (k)\Bigr]\,\,,
\label{D}
\end{equation}
where $f_B(k^0)$ is the boson occupation number
\begin{equation}
f_B(k^0)={1\over e^{|k^0|/T}-1}.
\end{equation}

\section{Thermal self-energy}
\label{3}
The neutrino self-energy operator $\Sigma(p)$ and the  
invariant amplitude for the $\nu_\ell\rightarrow~\nu_\ell$ transition are related by
\begin{equation}
{\cal{M}}(\nu_\ell\rightarrow\nu_\ell)=\bar{\nu}(p)\Sigma(p)\nu(p),
\label{M}
\end{equation}
and, in the Feynman gauge, the finite temperature self-energy operator is the sum of three 
diagrams, a bubble diagram with the $W$-boson, a bubble diagram with the scalar, and a tadpole diagram with the $Z$-boson
\begin{equation}
\Sigma(p)=\Sigma_W(p)+\Sigma_\Phi(p)+\Sigma_Z(p).
\label{Sigma}
\end{equation}
In the framework of the minimally extended electroweak standard model with an $SU(2)$-singlet right-handed neutrino, the two bubble diagrams can be written as \cite{Erdas:1990gy,Kuznetsov:2006ix}
\begin{equation}
\Sigma_W(p)=i{g^2\over 2}R{\gamma}_{\alpha}
\int{d^4k\over (2\pi)^4}S(k)G^{\alpha\beta}(p-k){\gamma}_{\beta}L,
\label{sigmaw}
\end{equation}
\begin{equation}
\Sigma_\Phi(p)=i{g^2\over 2m^2_W}[m_\ell R-m_\nu L]
\int{d^4k\over (2\pi)^4}S(k)D(p-k)[m_\ell L-m_\nu R],
\label{sigmaphi}
\end{equation}
where $g$ is the $SU(2)$ coupling constant, $L={1\over 2}(1-\gamma_5)$ and
$R={1\over 2}(1+\gamma_5)$ are the left-handed and right-handed projectors
and neutrino mixing is allowed by taking a nondiagonal neutrino mass matrix
$m_\nu$ in Eq. (\ref{sigmaphi}). The tadpole diagram calculation is easily found in the literature and does not need to be revisited in this paper, since it only contributes a local term to the self-energy that has been calculated exactly decades ago, unlike the bubble diagrams that contain non local terms. 

The bubble diagrams contain a vacuum part and two thermal parts due to contributions of 
thermal fermions, and thermal bosons respectively
\begin{equation}
\Sigma_{W}(p)=\Sigma_{W}^0(p)+
\Sigma_{W}^F(p)+\Sigma_{W}^B(p)\,\,,
\label{sigmaw2}
\end{equation}
\begin{equation}
\Sigma_{\Phi}(p)=\Sigma_{\Phi}^0(p)+
\Sigma_{\Phi}^F(p)+\Sigma_{\Phi}^B(p)\,\,.
\label{sigmaphi2}
\end{equation}
Inserting the expression for the vacuum propagators from Eqs. (\ref{S0}), (\ref{G0}), and (\ref{D0}) 
into the self-energy, I write ${\Sigma}_{W}^0(p)$ and ${\Sigma}_{\Phi}^0(p)$ as
\begin{equation}
\Sigma_{W}^0(p)=-{ig^2\over 2}
\int\!{{d^4 k}\over{{(2{\pi})}^4}}{\int}^{\infty}_{0}
{ds_1}{\int}^{\infty}_{0}
{ds_2}\,{e^{-is_1(m^2_\ell +k^2)}}
{e^{-is_2(m_W^2 +q^2)}}R{\gamma}_{\alpha}
(m_\ell-
\not\! k)
g^{\alpha\beta}{\gamma}_{\beta}L,
\end{equation}
\begin{equation}
\Sigma_{\Phi}^0(p)=-i{g^2\over 2m^2_W}
\int\!{{d^4 k}\over{{(2{\pi})}^4}}{\int}^{\infty}_{0}
{ds_1}{\int}^{\infty}_{0}
{ds_2}\,{e^{-is_1(m^2_\ell +k^2)}}
{e^{-is_2(m_W^2 +q^2)}}
[m_\ell R-m_\nu L](m_\ell-
\not\! k)
[m_\ell L-m_\nu R],
\end{equation}
where
\begin{equation}
q = p-k .
\end{equation}
Next I use the thermal propagators of Eqs. (\ref{S}), (\ref{G}), and (\ref{D}) and write the two thermal parts of each of the bubble diagrams
\begin{equation}
\Sigma_{W}^F(p)={ig^2\over 2}
\int\!{{d^4 k}\over{{(2{\pi})}^4}}
{\int}^{\infty}_{0}{ds_2}\,{e^{-is_2(m_W^2 +q^2)}}f_F(k^0)
{\int}^{\infty}_{0}{ds_1}\left[{e^{-is_1(m^2_\ell +k^2)}}+{e^{is_1(m^2_\ell +k^2)}}\right]
R{\gamma}_{\alpha}(m_\ell-\not\! k){\gamma}^{\alpha}L,
\label{sigmawF}
\end{equation}
\begin{equation}
\Sigma_{W}^B(p)=-{ig^2\over 2}
\int\!{{d^4 k}\over{{(2{\pi})}^4}}
{\int}^{\infty}_{0}{ds_1}{e^{-is_1(m^2_\ell +k^2)}}
f_B(k^0)
{\int}^{\infty}_{0}{ds_2}\left[{e^{-is_2(m_W^2 +q^2)}+e^{is_2(m_W^2 +q^2)}}\right]
R{\gamma}_{\alpha}(m_\ell-\not\! k){\gamma}^{\alpha}L,
\label{sigmawB}
\end{equation}
and
\begin{eqnarray}
\Sigma_{\Phi}^F(p)&=&{ig^2\over 2m^2_W}
\int\!{{d^4 k}\over{{(2{\pi})}^4}}
{\int}^{\infty}_{0}{ds_2}\,{e^{-is_2(m_W^2 +q^2)}}f_F(k^0)
{\int}^{\infty}_{0}{ds_1}\left[{e^{-is_1(m^2_\ell +k^2)}}+{e^{is_1(m^2_\ell +k^2)}}\right]
\nonumber \\
&&\times
(m_\ell R-m_\nu L)(m_\ell-\not\! k)(m_\ell L-m_\nu R),
\label{sigmaphiF}
\end{eqnarray}
\begin{eqnarray}
\Sigma_{\Phi}^B(p)&=&-{ig^2\over 2m^2_W}
\int\!{{d^4 k}\over{{(2{\pi})}^4}}
{\int}^{\infty}_{0}{ds_1}{e^{-is_1(m^2_\ell +k^2)}}
f_B(k^0)
{\int}^{\infty}_{0}{ds_2}\left[{e^{-is_2(m_W^2 +q^2)}+e^{is_2(m_W^2 +q^2)}}\right]
\nonumber \\
&&\times
(m_\ell R-m_\nu L)(m_\ell-
\not\! k)
(m_\ell L-m_\nu R).
\label{sigmaphiB}
\end{eqnarray}
Now I focus on $\Sigma_{W}^F(p)$, do the straightforward gamma algebra, change integration variable from $\vec{ k}$ to 
$\vec{ k}+{s_2\over{s_2+s_1}}\vec{ p}$ in the part that contains $e^{-is_1(m^2_\ell +k^2)}$, and change integration variable from $\vec{ k}$ to 
$\vec{ k}+{s_2\over{s_2-s_1}}\vec{ p}$ in the part that contains $e^{is_1(m^2_\ell +k^2)}$, and obtain
\begin{eqnarray}
\Sigma_{W}^F(p)&=&{-ig^2}
\int\!{{d^4 k}\over{{(2{\pi})}^4}}
{\int}^{\infty}_{0}{ds_2}\,{e^{-is_2m_W^2}}f_F(k^0)
{\int}^{\infty}_{0}{ds_1}
\nonumber \\
&&\times\left\{\exp\left[{-i\left(s_1m^2_\ell +(s_1+s_2)k^2-s_2(p^0)^2+2s_2k^0p^0+{s_1s_2\over{s_2+s_1}}\vec{p}\,^2\right)}\right]\right.
R(\not\! k+{s_2\over{s_2+s_1}}\not\!\vec{ p})L
\nonumber \\
&&+\left.\exp\left[{i\left(s_1m^2_\ell +(s_1-s_2)k^2+s_2(p^0)^2-2s_2k^0p^0+{s_1s_2\over{s_2-s_1}}\vec{p}\,^2\right)}\right]R(\not\! k+{s_2\over{s_2-s_1}}\not\!\vec{ p})L\right\},
\end{eqnarray}
where I use the standard notation $p^\mu = (p^0, \vec{p})$ and $k^\mu = (k^0, \vec{k})$. At this point I change integration variables from $s_1$ and $s_2$ to $s$ and $u$, with $s_1=su$ and $s_2=s(1-u)$, and obtain
\begin{eqnarray}
\Sigma_{W}^F(p)&=&{-ig^2}
\int\!{{d^4 k}\over{{(2{\pi})}^4}}
{\int}^{\infty}_{0}{sds}{\int}^{1}_{0}{du}\,{e^{-is(1-u)m_W^2}}f_F(k^0)
\nonumber \\
&&\times
\left\{\exp\left[{-is\left(um^2_\ell +k^2-(1-u)(p^0)^2+2(1-u)k^0p^0+u(1-u)\vec{p}\,^2\right)}\right]\right.
\nonumber \\
&&+\left.
\exp\left[{is\left(um^2_\ell +k^2+(1-u)(p^0)^2-2(1-u)k^0p^0+u(1-u)\vec{p}\,^2\right)}\right]\right\}R(\not\! k+(1-u)\not\!\vec{ p})L.
\end{eqnarray}
Next, I integrate over $d^3 \vec{k}$ and $s$, and find
\begin{equation}
\Sigma_{W}^F(p)={g^2}
\int^{\infty}_{-\infty}\!{{d k^0}\over{{(4{\pi})}^2}}f_F(k^0)
{\int}^{1}_{0}{du}\,(A_--A_+)
R(\gamma^0k^0-(1-u)\vec{\gamma}\cdot\vec{ p})L,
\label{sigmawF2}
\end{equation}
where $\gamma^\mu = (\gamma^0, \vec{\gamma})$ and I define
\begin{equation}
A_\pm=\left\{u\left[(k^0)^2-m^2_\ell\right]\pm(1-u)\left[m_W^2-(k^0)^2-(p^0)^2+2k^0p^0\right]-u(1-u)\vec{p}\,^2\right\}^{-1/2}.
\label{A}
\end{equation}
I do the $u$-integration and obtain the main result of this paper
\begin{equation}
\Sigma_{W}^F(p)={g^2}\!
\int^{\infty}_{-\infty}\!{{d k^0}\over{{(4{\pi})}^2}}{f_F(k^0)\over \left[m_W^2-(k^0)^2-(p^0)^2+2k^0p^0\right]^{1/2}}
\left\{[{\cal I}_1(x,y)-{\cal I}_2(x,y)]k^0
R\gamma^0L+[{\cal J}_1(x,y)-{\cal J}_2(x,y)]R\vec{\gamma}\cdot\vec{ p}L\right\},
\label{sigmawF3}
\end{equation}
where I introduce the dimensionless variables
\begin{equation}
x={(k^0)^2-m^2_\ell\over m_W^2-(k^0)^2-(p^0)^2+2k^0p^0},
\label{x}
\end{equation}
\begin{equation}
y={\vec{p}\,^2\over m_W^2-(k^0)^2-(p^0)^2+2k^0p^0},
\label{y}
\end{equation}
I define the two functions ${\cal I}_1(x,y)$ and ${\cal I}_2(x,y)$ to be
\begin{equation}
{\cal I}_1(x,y)={1\over\sqrt{y}}\ln\left({1+x+y+2\sqrt{x}\sqrt{y}}\over 1+x +y-2\sqrt{y}\right),
\label{I1}
\end{equation}
\begin{equation}
{\cal I}_2(x,y)={1\over\sqrt{y}}\ln\left({1-x-y-2\sqrt{x}\sqrt{y}}\over  1-x -y-2\sqrt{y}\right),
\label{I2}
\end{equation}
and I define the two functions ${\cal J}_1(x,y)$ and ${\cal J}_2(x,y)$ by means of integrals, since it is not possible to obtain a closed form of the integration 
\begin{equation}
{\cal J}_1(x,y)={\int}^{1}_{0}{du}(1-u)[ux-1+u-u(1-u)y]^{-1/2},
\label{J1}
\end{equation}
\begin{equation}
{\cal J}_2(x,y)={\int}^{1}_{0}{du}(1-u)[ux+1-u-u(1-u)y]^{-1/2}.
\label{J2}
\end{equation}
It is possible, however, to find asymptotic expansions of ${\cal J}_1(x,y)$ and ${\cal J}_2(x,y)$ in closed form. For $x\gg y$, I find
\begin{equation}
{\cal J}_1(x,0)-{\cal J}_2(x,0)={2\over 3}\left[{1-2\sqrt{x}\over(1-\sqrt{x})^2}-{1+2\sqrt{x}\over(1+\sqrt{x})^2}\right],
\label{J1-J2}
\end{equation}
and for $y\gg x$, I find
\begin{equation}
{\cal J}_1(0,y)={1\over y}-{(1+y)\arcsin \left(\sqrt{y\over 1+y}\right)\over y^{3/2}}
\label{J10y}
\end{equation}
\begin{equation}
{\cal J}_2(0,y)={1\over y}+{y-1\over 2y^{3/2}}\left[\pi-2\,\text{arcsec}\left(\sqrt{y-1\over y}\right)\right].
\label{J20y}
\end{equation}

Using the same method and similar steps, I obtain $\Sigma_{\Phi}^F(p)$
\begin{eqnarray}
\Sigma_{\Phi}^F(p)&=&{g^2\over 2m^2_W}
\int^{\infty}_{-\infty}\!{{d k^0}\over{{(4{\pi})}^2}}{f_F(k^0)\over \left[m_W^2-(k^0)^2-(p^0)^2+2k^0p^0\right]^{1/2}}
\nonumber \\
&&\times\left\{[{\cal I}_1(x,y)-{\cal I}_2(x,y)]k^0
{\cal R}\gamma^0{\cal L}+[{\cal J}_1(x,y)-{\cal J}_2(x,y)]{\cal R}\vec{\gamma}\cdot\vec{ p}{\cal L}\right\},
\label{sigmaphiF2}
\end{eqnarray}
where I use the notation ${\cal R}=[m_\ell R-m_\nu L]$ and ${\cal L}=[m_\ell L-m_\nu R]$. It is evident that $\Sigma_{\Phi}^F(p)$ is similar to $\Sigma_{W}^F(p)$, but it is suppressed by a factor of $m^2_\ell\over m^2_W$, or ${m_\ell m_\nu}\over m^2_W$, or $m^2_\nu\over m^2_W$, and therefore can be safely neglected when compared to $\Sigma_{W}^F(p)$.

Next I evaluate $\Sigma_{W}^B(p)$ and obtain
\begin{equation}
\Sigma_{W}^B(p)=-{g^2}
\int^{\infty}_{-\infty}\!{{d k^0}\over{{(4{\pi})}^2}}f_B(k^0)
{\int}^{1}_{0}{du}\,(B_--B_+)
R(\gamma^0k^0-(1-u)\vec{\gamma}\cdot\vec{ p})L,
\label{sigmawB2}
\end{equation}
where I define
\begin{equation}
B_\pm=\left\{u\left[(k^0)^2-m^2_W\right]\pm(1-u)\left[m_\ell^2-(k^0)^2-(p^0)^2+2k^0p^0\right]-u(1-u)\vec{p}\,^2\right\}^{-1/2}.
\label{B}
\end{equation}
Notice the presence of the boson occupation number, $f_B(k^0)$, inside $\Sigma_{W}^B(p)$. Only if one investigates temperatures above 10 GeV the thermal excitations of the $W$ boson become relevant and one will have to consider $\Sigma_{W}^B(p)$. Therefore, in most cases, $\Sigma_{W}^B(p)$ can safely be neglected .

Finally, $\Sigma_{\Phi}^B(p)$ is similar to $\Sigma_{W}^B(p)$ but suppressed by a factor of  $m^2_\ell\over m^2_W$, or ${m_\ell m_\nu}\over m^2_W$, or $m^2_\nu\over m^2_W$, thus can be safely neglected. 

\section{Low temperature and energy}
\label{4}
In this section I will show that, at low temperature and energy and to lowest order in $1\over m_W$, the exact self-energy of Eq. (\ref{sigmawF3}) returns the well-known result obtained by the contact approximation. l will also obtain a simple expression for the next order corrections which are the leading terms for the self-energy in a $CP$-symmetric medium. 

When $T$, $\vec{p} \ll m_W$, values of $k^0$ much larger than $T$
are strongly suppressed by the fermion occupation number $f_F(k^0)$ and therefore one can also take $k^0 \ll m_W$, in addition to $p^0 \ll m_W$ which follows immediately from the fact that the neutrino momentum $\vec{p} \ll m_W$. Under these assumptions, the two dimensionless variables defined in Eqs. (\ref{x}) and (\ref{y}) become
 \begin{equation}
x\simeq{(k^0)^2-m^2_\ell\over m_W^2},
\label{xsmall}
\end{equation}
\begin{equation}
y={\vec{p}\,^2\over m_W^2},
\label{ysmall}
\end{equation}
and, for $x\ll 1$ and $y\ll 1$, I find
\begin{equation}
{\cal I}_1(x,y)-{\cal I}_2(x,y)\simeq 4\sqrt{x} - 4(x+y)+{\cal O}(x^{3/2}, y^{3/2}),
\label{I1-I2}
\end{equation}
\begin{equation}
{\cal J}_1(x,0)-{\cal J}_2(x,0)\simeq -{8\over 3}x^{3/2}+{\cal O}(x^{5/2}),
\label{J1-J2_x}
\end{equation}
\begin{equation}
{\cal J}_1(0,y)-{\cal J}_2(0,y)\simeq {4\over 15}y + {4\over 63}y^3 + {\cal O}(y^5).
\label{J1-J2_y}
\end{equation}
Substituting Eqs. (\ref{I1-I2}) and (\ref{J1-J2_x}) or (\ref{J1-J2_y}) into Eq. (\ref{sigmawF3}), using $\left[m_W^2-(k^0)^2-(p^0)^2+2k^0p^0\right]^{1/2}\simeq \,m_W$, and neglecting terms of order higher than $\left(1\over m_W\right)^2$, I find the value of the exact self-energy, $\Sigma_{W}^F(p)$, when $T$, $\vec{p} \ll m_W$
\begin{equation}
\Sigma_{W}^F(p)={g^2\over m_W^2}
\int^{\infty}_{-\infty}\!{{d k^0}\over{{(2{\pi})}^2}}f_F(k^0)
k^0\sqrt{(k^0)^2-m^2_\ell}\,\,
R\gamma^0L,
\label{sigmawFlow1}
\end{equation}
which, with the change of integration variable $k^0=\sqrt{k^2+m^2_\ell}$, becomes
\begin{equation}
\Sigma_{W}^F(p)={g^2\over m_W^2}
\int^{\infty}_0\!{k^2d k\over(2\pi)^2}
\left[f_F^+(\omega_k)-f_F^-(-\omega_k)\right]
R\gamma^0L,
\label{sigmawFlow2}
\end{equation}
where I use $\omega_k=\sqrt{k^2+m^2_\ell}$. Eq. (\ref{sigmawFlow2}) can be written as
\begin{equation}
\Sigma_{W}^F(p)={g^2\over 4m_W^2}
(n_\ell-n_{\bar{\ell}})
R\gamma^0L,
\label{sigmawFlow3}
\end{equation}
where $n_\ell-n_{\bar{\ell}}$ is the net lepton number density in the medium
\begin{equation}
n_\ell-n_{\bar{\ell}}=
2\int{d^3 k\over(2\pi)^3}
\left[f_F^+(\omega_k)-f_F^-(-\omega_k)\right].
\label{nell}
\end{equation}
Eq. (\ref{sigmawFlow3}) is the well known result for $\Sigma_{W}^F(p)$ evaluated in the contact approximation \cite{giuntibook,Erdas:1998uu,Bethe:1986ej,Notzold:1987ik,Pal:1989xs}, and can be written in a manifestly covariant form using the four velocity of the medium $u^\mu$ and setting $\not\!{ u} = -\gamma^0$ in the reference frame of the medium.

If the medium is $CP$-symmetric one needs to dig further, since $n_\ell-n_{\bar{\ell}}=0$. I will examine two asymptotic cases, $T\gg \vec{p}$ and $\vec{p}\gg T$, while maintaining $T$, $\vec{p} \ll m_W$. If $T\gg \vec{p}$, then $x\gg y$ and I use Eq. (\ref{J1-J2_x}) and Eq. (\ref{xsmall}), to find
\begin{equation}
\Sigma_{W}^F(p)=-{g^2\over m^4_W}
\int^{\infty}_{-\infty}\!{{d k^0}\over{6{\pi}^2}}{f_F(k^0)}[(k^0)^2-m^2_\ell]^{3/2}
R\vec{\gamma}\cdot\vec{ p}L.
\label{sigmaCP1}
\end{equation}
Notice that, in the $CP$-symmetric medium, the integral multiplying $R\gamma^0L$ vanishes since the integrand is an odd function of $k^0$. Changing the integration variable from $k^0$ to $\omega_k$ as defined above, I find
\begin{equation}
\Sigma_{W}^F(p)=-{g^2\over m^4_W}
\int^{\infty}_0\!{{k^2d k}\over{6{\pi}^2}}{k^2\over \omega_k}\left[f_F^+(\omega_k)+f_F^-(-\omega_k)\right]
R\vec{\gamma}\cdot\vec{ p}L,
\label{sigmaCP2}
\end{equation}
which can be written in manifestly covariant form using $\vec{\gamma}\cdot\vec{ p}\simeq \gamma^0 p^0 + \not\!{ p}=
(u\cdot p)\not\!{u} \,+ \not\!{ p}$ in the reference frame of the medium. The remaining integral can be evaluated, and its value is
\begin{equation}
\int^{\infty}_0\!{d k}{k^4\over \omega_k}\left[f_F^+(\omega_k)+f_F^-(-\omega_k)\right]={7\pi^4\over 60}T^4\left[1+{\cal O}(m^2_\ell/T^2)\right],
\label{Int1}
\end{equation}
which yields
\begin{equation}
\Sigma_{W}^F(p)=-{g^2}{7\pi^2\over 360}{T^4\over m^4_W}
R\vec{\gamma}\cdot\vec{ p}L.
\label{sigmaCP2a}
\end{equation}

When $\vec{p}\gg T$ it is also $y\gg x$, and I use Eq. (\ref{J1-J2_y}) and Eq. (\ref{xsmall}) to find
\begin{equation}
\Sigma_{W}^F(p)={g^2\over m^3_W}
\int^{\infty}_{-\infty}\!{{d k^0}\over{60{\pi}^2}}{f_F(k^0)}\vec{p}\,^2
R\vec{\gamma}\cdot\vec{ p}L.
\label{sigmaCP3}
\end{equation}
which, with the change of integration variable described above, becomes
\begin{equation}
\Sigma_{W}^F(p)={g^2\over m^3_W}
\int^{\infty}_0\!{{kd k}\over{60{\pi}^2}}{\vec{p}\,^2\over \omega_k}\left[f_F^+(\omega_k)+f_F^-(-\omega_k)\right]
R\vec{\gamma}\cdot\vec{ p}L,
\label{sigmaCP4}
\end{equation}
and I write it in covariant form using $\vec{p}\,^2=(p^0)^2+p^2=(u\cdot p)^2 +p^2$ in the frame of the medium, and the expression introduced below Eq. (\ref{sigmaCP2}) for $\vec{\gamma}\cdot\vec{ p}$. I evaluate the remaining integral
\begin{equation}
\int^{\infty}_0\!{d k}{k\over \omega_k}\left[f_F^+(\omega_k)+f_F^-(-\omega_k)\right]=2T\ln(2)\left[1+{\cal O}(m^2_\ell/T^2)\right],
\label{Int2}
\end{equation}
and obtain
\begin{equation}
\Sigma_{W}^F(p)={g^2}{\ln(2)\over{30{\pi}^2}}
{T\vec{p}\,^2\over m^3_W}
R\vec{\gamma}\cdot\vec{ p}L,
\label{sigmaCP4a}
\end{equation}

Notice that, if $\vec{p}\gg T$, $\Sigma_{W}^F(p)$ is suppressed by $1\over m^3_W$, while, if $T\gg \vec{p}$, it is suppressed by $1\over m^4_W$ and thus much smaller.

\section{High temperature and energy}
\label{5}
I will examine two scenarios in the high temperature and energy regime: $ \vec{p}\sim m_W \gg T$, and $T\sim m_W \gg \vec{p}$. These are two of the most relevant high temperature and energy situations since, when $T\gg m_W $ the $SU(2)\otimes U(1)$ symmetry is restored and the $W$-boson is massless, requiring a different calculation than the one presented in this paper. We can also safely assume that at such temperatures the medium is  $CP$-symmetric.

In the first case,  $ \vec{p}\sim m_W \gg T$, I find that the two dimensionless variables $x$ and $y$ are:
\begin{equation}
x\simeq{(k^0)^2-m^2_\ell\over m_W^2-(p^0)^2},
\label{x1}
\end{equation}
\begin{equation}
y\simeq{\vec{p}\,^2\over m_W^2-(p^0)^2},
\label{y1}
\end{equation}
since $p^0\sim \vec{p}$ and therefore $(p^0-k^0)^2\simeq (p^0)^2$, because values of $k^0$ much larger than the temperature $T$ are suppressed by the Fermi-Dirac distribution. It is evident that, under these conditions, $y\gg x$ and $y\gg1$. Under these assumptions, and that of a $CP$-symmetric medium, the integral multiplying $R\gamma^0L$ vanishes in the exact self-energy of Eq. (\ref{sigmawF3}) and I can use Eqs. (\ref{J10y}) - (\ref{J20y}) for ${\cal J}_1$ and ${\cal J}_2$, to find
\begin{equation}
\Sigma_{W}^F(p)={g^2}
\int^{\infty}_{-\infty}\!{{d k^0}\over{16{\pi}}}{f_F(k^0)\over \vec{p}}
R\vec{\gamma}\cdot\vec{ p}L,
\label{sigmahp1}
\end{equation}
where I used
\begin{equation}
{\cal J}_1(0,y)-{\cal J}_2(0,y)\simeq {\pi\over \sqrt{y}} + {\cal O}(1/y),
\label{J1-J2_ly}
\end{equation}
for $y\gg 1$. Eq. (\ref{sigmahp1}), after the change of integration variable used in the previous section, can be written in a more familiar form
\begin{equation}
\Sigma_{W}^F(p)={g^2}
\int^{\infty}_0\!{{d k}\over{16{\pi}}}{k\over \omega_k}\left[f_F^+(\omega_k)+f_F^-(-\omega_k)\right]
R\vec{\gamma}\cdot\hat{p}L,
\label{sigmahp1a}
\end{equation}
where $\hat{p}$ is a unit vector in the direction of the neutrino momentum. Using Eq. (\ref{Int2}) to evaluate the integral, I find
\begin{equation}
\Sigma_{W}^F(p)={g^2}
{{\ln(2)}\over{8{\pi}}}T
R\vec{\gamma}\cdot\hat{p}L.
\label{sigmahp1b}
\end{equation}

In the second case, $ T\sim m_W \gg \vec{p}$, I find
\begin{equation}
x\simeq{(k^0)^2-m^2_\ell\over m_W^2-(k^0)^2},
\label{x2}
\end{equation}
\begin{equation}
y\simeq{\vec{p}\,^2\over m_W^2-(k^0)^2},
\label{y2}
\end{equation}
with $x\gg y$ and $x\gg 1$, since I can take $k^0\sim T\gg p^0$. I evaluate the self-energy in this scenario, using Eq. (\ref{sigmawF3}), and again find that the integral multiplying $R\gamma^0L$ vanishes because the medium is $CP$ symmetric.
I will use Eq. (\ref{J1-J2}) for ${\cal J}_1 - {\cal J}_2$, to find
\begin{equation}
\Sigma_{W}^F(p)={g^2}
\int^{\infty}_{-\infty}\!{{d k^0}\over{6{\pi}^2}}{f_F(k^0)\over \sqrt{(k^0)^2-m^2_\ell}}
R\vec{\gamma}\cdot\vec{ p}L,
\label{sigmahp2}
\end{equation}
where I used
\begin{equation}
{\cal J}_1(x,0)-{\cal J}_2(x,0)\simeq {8\over 3 \sqrt{x}} + {\cal O}(1/x),
\label{J1-J2_hx}
\end{equation}
for $x\gg 1$. After the change of integration variable used previously in this section, I obtain 
\begin{equation}
\Sigma_{W}^F(p)={g^2}
\int^{\infty}_0\!{{d k}\over{6{\pi}^2}}{\left[f_F^+(\omega_k)+f_F^-(-\omega_k)\right]\over \omega_k}
R\vec{\gamma}\cdot\vec{ p}L,
\label{sigmahp3}
\end{equation}
and evaluate the remaining integral
\begin{equation}
\int^{\infty}_0{dk\over \omega_k}\left[f_F^+(\omega_k)+f_F^-(-\omega_k)\right]={1\over 2}\ln\left({T\over m_\ell}\right)\left[1+{\cal O}(m^2_\ell/T^2)\right],
\label{Int3}
\end{equation}
to find
\begin{equation}
\Sigma_{W}^F(p)={g^2\over{12{\pi}^2}}\ln\left({T\over m_\ell}\right)
R\vec{\gamma}\cdot\vec{ p}L.
\label{sigmahp4}
\end{equation}
\section{Conclusions}
\label{6}
In this work I used Schwinger's proper time method to calculate the exact neutrino self-energy bubble diagrams at finite temperature and density, without resorting to the contact approximation that has always been used to approximately accomplish this task \cite{Wolfenstein:1977ue,Mikheev:1986gs,Bethe:1986ej,Notzold:1987ik,Pal:1989xs,Erdas:1998uu}. At low temperature, $T\ll m_e$, the self-energy I derive reduces to the well established form obtained with the contact approximation. However, at higher temperature, non-local terms in the self-energy become more relevant and cannot be neglected when evaluating the dispersion of neutrinos in a hot medium. My work, by evaluating the self-energy exactly, accurately retains and keeps track of all local and non-local terms in the thermal self-energy.

Using my main result of Eq. (\ref{sigmawF3}), I obtain simple analytic forms of the self-energy in a variety of scenarios that explore several combinations of the following conditions: $CP$-asymmetric and $CP$-symmetric plasma, low and high temperature, low and high neutrino energy. In all cases I obtained the expected covariant form of the self-energy
\begin{equation}
\Sigma=R(a \not\!{p} \,+ b \not\!{ u})L,
\label{sigma_gen}
\end{equation}
where $p^\mu$ is the neutrino momentum and $u^\mu$ is the four velocity of the medium. I find the neutrino dispersion relation in the medium by setting $(\not\!{p}+a \not\!{p} \,+ b \not\!{ u})^2=0$ and neglecting the small neutrino mass. To lowest order, I obtain the following 
\begin{equation}
E =\vert \vec{p}\vert -b,
\label{disp_1}
\end{equation}
where the neutrino 4-momentum is $p^\mu = (E,\vec{p})$ and $-b$ can be interpreted as a neutrino effective mass.

In a $CP$-asymmetric medium at low temperature, I use Eq. (\ref{sigmawFlow3}) and obtain the well known result \cite{Wolfenstein:1977ue,Wolfenstein:1979ni,Mikheev:1986gs,Bethe:1986ej,Notzold:1987ik,Pal:1989xs,Erdas:1998uu}
\begin{equation}
b=-{g^2\over 4m_W^2}(1+c_V)
(n_\ell-n_{\bar{\ell}}),
\label{b1}
\end{equation}
where I added the tadpole diagram contribution, proportional to $c_V=-{1\over 2}+2\sin^2\theta_W$ where $\theta_W$ is the Weinberg angle.

When the medium is $CP$-symmetric and temperature and energy are low, with $T \gg E$, I obtain from Eq. (\ref{sigmaCP2a})
\begin{equation}
b={g^2}{7\pi^2\over 360}{T^4\over m^4_W}E,
\label{b2}
\end{equation}
when $E \gg T$, I use Eq. (\ref{sigmaCP4a}) to find
\begin{equation}
b=-{g^2}{\ln(2)\over{30{\pi}^2}}
{T\vec{p}\,^2\over m^3_W}E.
\label{b3}
\end{equation}

In the case of a $CP$-symmetric medium and high neutrino energy, with $E \sim m_W \gg T$, I use Eq. (\ref{sigmahp1b}) and find
\begin{equation}
b=-{g^2}
{{\ln(2)}\over{8{\pi}}}{T\over \vert \vec{p}\vert}E,
\label{b4}
\end{equation}
while, when $T \sim m_W \gg E$, I use Eq. (\ref{sigmahp4}) to obtain
\begin{equation}
b=-
{g^2\over{12{\pi}^2}}\ln\left({T\over m_\ell}\right)E.
\label{b5}
\end{equation}


\begin{thebibliography}{99}

\bibitem{raffeltbook}
G.~G.~Raffelt,
{\em Stars as Laboratories for Fundamental Physics}
(University of Chicago Press, Chicago, 1996).

\bibitem{giuntibook}
C.~Giunti and C.~W.~Kim,
{\em Fundamentals of Neutrino Physics and Astrophysics}
(Oxford, UK: University Press, 2007).

\bibitem{Wolfenstein:1977ue} 
  L.~Wolfenstein,
  Phys.\ Rev.\ D {\bf 17}, 2369 (1978).

 \bibitem{Wolfenstein:1979ni} 
  L.~Wolfenstein,
  Phys.\ Rev.\ D {\bf 20}, 2634 (1979).
  
\bibitem{Langacker:1982ih}
P.~Langacker, J.~P.~Leveille and J.~Sheiman,
Phys. Rev. D \textbf{27}, 1228 (1983)

\bibitem{Mikheev:1986gs} 
  S.~P.~Mikheyev and A.~Y.~Smirnov,
  Sov.\ J.\ Nucl.\ Phys.\  {\bf 42}, 913 (1985)
  [Yad.\ Fiz.\  {\bf 42}, 1441 (1985)].

 \bibitem{Bethe:1986ej} 
  H.~A.~Bethe,
  Phys.\ Rev.\ Lett.\  {\bf 56}, 1305 (1986).
  
\bibitem{Notzold:1987ik}
D.~Notzold and G.~Raffelt,
Nucl. Phys. B \textbf{307}, 924-936 (1988).

\bibitem{Pal:1989xs}
P.~B.~Pal and T.~N.~Pham,
Phys. Rev. D \textbf{40}, 259 (1989).

\bibitem{Enqvist:1990ad}
K.~Enqvist, K.~Kainulainen and J.~Maalampi,
Nucl. Phys. B \textbf{349}, 754-790 (1991)

\bibitem{Langacker:1992xk}
P.~Langacker and J.~Liu,
Phys. Rev. D \textbf{46}, 4140-4160 (1992)

\bibitem{DOlivo:1992lwg}
J.~C.~D'Olivo, J.~F.~Nieves and M.~Torres,
Phys. Rev. D \textbf{46}, 1172-1179 (1992)

\bibitem{Friedland:2003dv}
A.~Friedland and C.~Lunardini,
Phys. Rev. D \textbf{68}, 013007 (2003)

\bibitem{Erdas:1990gy}
A.~Erdas and G.~Feldman,
Nucl. Phys. B \textbf{343}, 597-621 (1990).

\bibitem{Erdas:1998uu}
A.~Erdas, C.~W.~Kim and T.~H.~Lee,
Phys. Rev. D \textbf{58}, 085016 (1998).

\bibitem{Erdas:2009zh}
A.~Erdas,
Phys. Rev. D \textbf{80}, 113004 (2009).

\bibitem{Borisov:1985ha}
A.~V.~Borisov, V.~C.~Zhukovsky, A.~V.~Kurilin and A.~I.~Ternov,
Yad. Fiz. \textbf{41}, 743-748 (1985).

\bibitem{Borisov:1988wy}
A.~V.~Borisov, V.~C.~Zhukovsky and A.~I.~Ternov,
Sov. Phys. J. \textbf{31}, 228-233 (1988)

\bibitem{Borisov:1989yw}
A.~V.~Borisov, V.~C.~Zhukovsky and A.~I.~Ternov,
Sov. Phys. Dokl. \textbf{34}, 884-885 (1989)

\bibitem{Elmfors:1996gy}
P.~Elmfors, D.~Grasso and G.~Raffelt,
Nucl. Phys. B \textbf{479}, 3-24 (1996).

\bibitem{DOlivo:1989ued}
J.~C.~D'Olivo, J.~F.~Nieves and P.~B.~Pal,
Phys. Rev. D \textbf{40}, 3679 (1989).

\bibitem{Kuznetsov:2005tq}
A.~V.~Kuznetsov, N.~V.~Mikheev, G.~G.~Raffelt and L.~A.~Vassilevskaya,
Phys. Rev. D \textbf{73}, 023001 (2006)
[erratum: Phys. Rev. D \textbf{73}, 029903 (2006)].

\bibitem{Mckeon:1981ym}
G.~Mckeon,
Phys. Rev. D \textbf{24}, 2744-2747 (1981)

\bibitem{Elizalde:2000vz}
E.~Elizalde, E.~J.~Ferrer and V.~de la Incera,
Annals Phys. \textbf{295}, 33-49 (2002).

\bibitem{Elizalde:2004mw}
E.~Elizalde, E.~J.~Ferrer and V.~de la Incera,
Phys. Rev. D \textbf{70}, 043012 (2004).

\bibitem{Nieves:2018qwg} 
  J.~F.~Nieves and S.~Sahu,
Eur.\ Phys.\ J.\ C {\bf 78}, no. 7, 547 (2018).

\bibitem{Tarazona:2017jnd} 
  C.~G.~Tarazona, A.~Castillo, R.~A.~Diaz and J.~Morales,
  arXiv:1706.08614 [hep-ph].

\bibitem{Popov:2019nkr} 
  A.~Popov and A.~Studenikin,
  Eur.\ Phys.\ J.\ C {\bf 79}, no. 2, 144 (2019).

\bibitem{Kuznetsov:2006ix}
A.~V.~Kuznetsov and N.~V.~Mikheev,
[arXiv:hep-ph/0605114 [hep-ph]].

 \bibitem{Schwinger:1951nm}
J.~S.~Schwinger,
Phys. Rev. \textbf{82}, 664-679 (1951).

\end{thebibliography}
\end{document}